\documentclass[aps,prl,twocolumn,superscriptaddress]{revtex4-1}

\usepackage{graphicx}
\usepackage{amssymb}

\begin{document}
\title{Majorana modes and  $p$-wave superfluids for fermionic atoms in optical lattices}

\author{A.~B\"uhler}
\affiliation{Institute for Theoretical Physics III, University of Stuttgart, D-70550 Stuttgart, Germany}
\author{N. Lang}
\affiliation{Institute for Theoretical Physics III, University of Stuttgart, D-70550 Stuttgart, Germany}
\author{C.V. Kraus}
\affiliation{Institute for Quantum Optics and Quantum Information of the Austrian Academy of Sciences, A-6020 Innsbruck, Austria} 
\affiliation{Institute for Theoretical Physics, Innsbruck University, A-6020 Innsbruck, Austria}
\author{G. M\"oller}
\affiliation{TCM Group, Cavendish Laboratory, J.J. Thomson Avenue, Cambridge CB3 0HE, UK}
\author{S.D. Huber}
\affiliation{Institute for Theoretical Physics, ETH Zurich, 8093 Z\"urich, Switzerland}
\author{H.P. B\"uchler}
\affiliation{Institute for Theoretical Physics III, University of Stuttgart, D-70550 Stuttgart, Germany}

\date{\today}

\begin{abstract}
 We present a simple approach to create a strong $p$-wave interaction for fermions in an optical lattice. 
 The crucial step is that the combination of a lattice setup with different orbital states and $s$-wave interactions can give rise to a 
 strong induced $p$-wave pairing. We identify different topological phases and demonstrate that the setup offers a natural way to explore
  the  transition from Kitaev's Majorana wires to two-dimensional $p$-wave superfluids. We demonstrate how this design can induce Majorana modes at edge dislocations in the optical lattice, and we provide an experimentally feasible protocol for the observation of the non-Abelian statistics.
\end{abstract}

\maketitle

\section{Introduction}

The quest for realisations of non-Abelian phases of matter, driven by their possible use in fault-tolerant 
topological quantum computing, has been spearheaded by recent developments in $p$-wave superconductors. The chiral $p_x + i p_y$-wave superconductor in two-dimensions exhibiting Majorana modes provides the simplest phase supporting non-Abelian quasiparticles  and can be seen as the blueprint of fractional topological order.  Alternatively, Kitaev's Majorana wire has emerged as an ideal toy model to understand Majorana modes. Here, we present a way to make the transition from Kitaev's Majorana wires to two-dimensional $p$-wave superconductors in a system with cold atomic gases in an optical lattice. The main idea is based on an approach to generate $p$-wave interactions by coupling orbital degrees of freedom  with strong $s$-wave interactions. We demonstrate how this design can induce Majorana modes at edge dislocations in the optical lattice and we provide an experimentally feasible protocol for the observation of the non-Abelian statistics.

Candidates for topological phases supporting non-Abelian anyons   \cite{Moore:1991vr}  with potential application in topological quantum computing \cite{Kitaev:2003jw,TQCReview} are found among a variety of systems including superfluid $^3$He-A \cite{Volovik+1992}, the layered superconductor Sr$_2$Ru\,O$_4$ \cite{Ishida:1998es}, the fractional quantum Hall state at $\nu=5/2$ \cite{Jiang:2011uy, Willett:2013fd}, and superconductor / topological insulator or similar heterostructures \cite{Sau:2010cl,Alicea:2010hy,Oreg2010,Beenakker2013}. Most recently, indium antimonide nano-wires in contact with an $s$-wave superconductor have shown promising experimental evidence consistent with the presence of the sought-after non-Abelian zero-energy Majorana states \cite{Mourik:2012je,Das2012}. However, many questions still ask for a definitive answer. 

Alongside the tremendous progress in solid-state systems, cold atomic gases provide a different angle when looking at $p$-wave superconductors. Thanks to their largely different strengths and shortcomings compared to solid-state systems, cold atomic gases might offer solutions to problems that are yet hard to address otherwise. For instance, it is well known that the spatial dimension of a setup can easily be controlled by optical lattices, while Feshbach resonances allow one to tune the interaction strength almost at will \cite{2007AnPhy.322....2G}. Unfortunately, the lifetime of $p$-wave resonant gases was found to be very limited \cite{Zhang:2004gq,Gaebler:2007jo} due to a number of well understood decay channels \cite{Levinsen:2008ia,JonaLasinio:2008il}. Identifying realisations of atomic $p$-wave superfluids with a sufficient lifetime emerged as a central challenge in this field. This led to proposals such as Bose-Fermi mixtures in shallow \cite{Wang:2005fi} or deep \cite{Massignan:2010hm} optical lattices, microwave dressed polar molecules \cite{Cooper:2009bj},  the introduction of synthetic spin-orbit coupling into an $s$-wave superfluid \cite{Sato:2009id},  the quantum Zeno effect \cite{Han:2009fk}, or driven dissipation \cite{Bardyn2012}.   However, the complexity in these proposed setups has so far precluded an experimental realisation.

\begin{figure}[t]
 \includegraphics[width= 0.8\columnwidth]{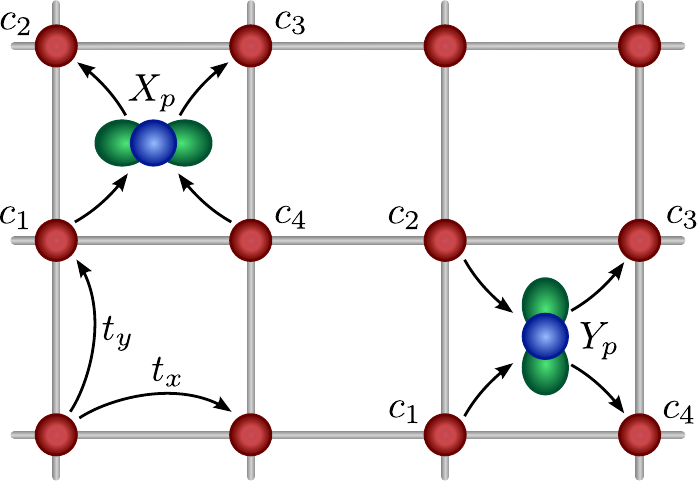}
  \caption{\textbf{Lattice setup.} 
    Spinless fermions residing at the lattice sites are coupled to a molecular state in the center of each plaquette. 
    The molecular states exhibit a $p$-wave symmetry and are doubly degenerate. Anisotropic hoppings $t_{y}/t_{x}$ 
    allow for the transition from coupled wires to the 2D isotropic system.
   }\label{fig1}
\end{figure}

Here, we present a simple approach to create a strong $p$-wave interaction for fermions in an optical lattice. 
The main idea is based on a resonant coupling from the lattice sites to a molecular state residing in the center of the 
plaquette in analogy to Ref.~\cite{buchler05,kraus13}. The crucial step is that the combination of a lattice setup with different orbital states and $s$-wave interactions can give rise to a strong induced $p$-wave pairing; similar ideas have recently been proposed \cite{Liu2014}.
We will demonstrate the appearance of $p$-wave superfluid phases via this coupling.
Moreover, a setup intrinsically based on an optical lattice allows one to naturally explore the transition from a two-dimensional $p$-wave superfluid to Kitaev's Majorana wire \cite{Kitaev:2001gb}. 
Hence we identify different topological transitions \cite{daichi12}, where the combination of the Fermi-surface topology with the symmetry of the $p$-wave superfluid order parameter gives rise to a rich phase diagram.  
Most remarkably, we find the appearance of Majorana modes localized at edge dislocations; such edge dislocations correspond to vortices in the phase of the lasers generating the lattice. In combination with another realistic ingredient to modern cold atoms experiments, single site addressability in the lattice \cite{bakr09,sherson10}, we provide a protocol for the observation of the non-Abelian braiding statistics.

\section{Effective Hamiltonian}

We start with the presentation of the Hamiltonian underlying our system. We focus on a setup of spinless fermionic atoms in a quadratic two-dimensional optical lattice. Then, the Hamiltonian is well described by the tight-binding model
\begin{equation}
  H =- \sum_{\langle  i  j\rangle} t_{i j} c^{\dag}_{i} c_{j}  -\mu \sum_{i} c^{\dag}_{i} c_{i}+ H_{x} +H_{y}. \label{hopping}
 \end{equation}
Here, $c_{i}^{\dag}$  and $c_{i}$ denote the fermionic creation (annihilation) operators at lattice site $i$, while  $\mu$ is the chemical potential fixing the average particle number, and  $t_{i j}$ denotes the hopping energy between nearest neighbor sites  $\langle  i j \rangle$.  In order to study the transition from a bulk two-dimensional setup to weakly coupled one-dimensional chains, we allow for an anisotropic hopping $t_{i j}$,  where $t_{ij}=t_{x(y)}$ for hopping along a link in the x-(y-)direction, respectively.
The interaction between the fermions is driven by resonant couplings $H_{x(y)}$ to two distinct lattice bound states $X_p$ and $Y_p$ residing in the center of each plaquette as shown in Fig.~\ref{fig1}; similar setups for bosonic atoms have been previously proposed  \cite{buchler05}.  
For spinless fermions on the lattice sites, these bound states must exhibit an odd parity symmetry for a non-vanishing interaction, which in our situation is a two-fold degenerate $p$-wave symmetry. Then, the coupling Hamiltonians reduce to
\begin{eqnarray}
  H_{x} & =& \gamma \sum_{p}X^{\dag}_{p} X_{p}  + g  \sum_{p} \left[ X^{\dag}_{p}  \left( c_{2} c_{3} - c_{4} c_{1}\right)+ {\rm h.c.} \right],  \nonumber \\
  H_{y}  &=& \gamma \sum_{p} Y^{\dag}_{p} Y_{p}  + g   \sum_{p} \left[ Y^{\dag}_{p}  \left( c_{1} c_{2} - c_{3} c_{4}\right)+ {\rm h.c.} \right]
  \label{molecularcoupling},
 \end{eqnarray}
where the summation $\sum_{p}$ runs over all plaquettes. The four lattice sites surrounding each plaquette are labelled as shown in  Fig.~\ref{fig1}.
The couplings to the lattice bound states respect the $p$-wave symmetry with coupling strength  $g$, while the detuning from resonance is given by $\gamma$. The latter quantity also includes the chemical potential $\gamma = \hbar \omega-  2 \mu$ with $\hbar\omega$ the energy difference between the molecular state and two free fermions. 

The most crucial part is the possibility to induce a strong $p$-wave interaction by the combination of orbital degrees of freedom and $s$-wave interactions. Here, we provide a sketch of this fundamental idea (for details we refer to the Appendix): The two-particle states $X_{p}$ and $Y_{p}$ in the center of the plaquette consist of two orbital states in the optical lattice forming a repulsively bound state \cite{winkler06}.  In order for these lattice molecules to fulfill the $p$-wave symmetry, we choose the lowest and the first excited state in the lattice confining the atoms in the center of the plaquette. Furthermore, the two fermions in the two orbital states have to be in different hyperfine states in order to profit from a stable $s$-wave interaction which can be tuned by conventional Feshbach resonances \cite{Chin2010}. This requires the coupling to the plaquette states to induce transitions between hyperfine states. To summarize: the $s$-wave interaction leads to the formation of repulsively bound pairs, while the orbital degree of freedom is responsible for the $p$-wave character of these lattice bound molecules. It is via this mechanism that the optical lattice breaks rotational symmetry and couples to states with different orbital symmetry  allowing for the conversion of $s$-wave to $p$-wave interactions.

\begin{figure*}[t]
  \includegraphics[width= 1.0\linewidth]{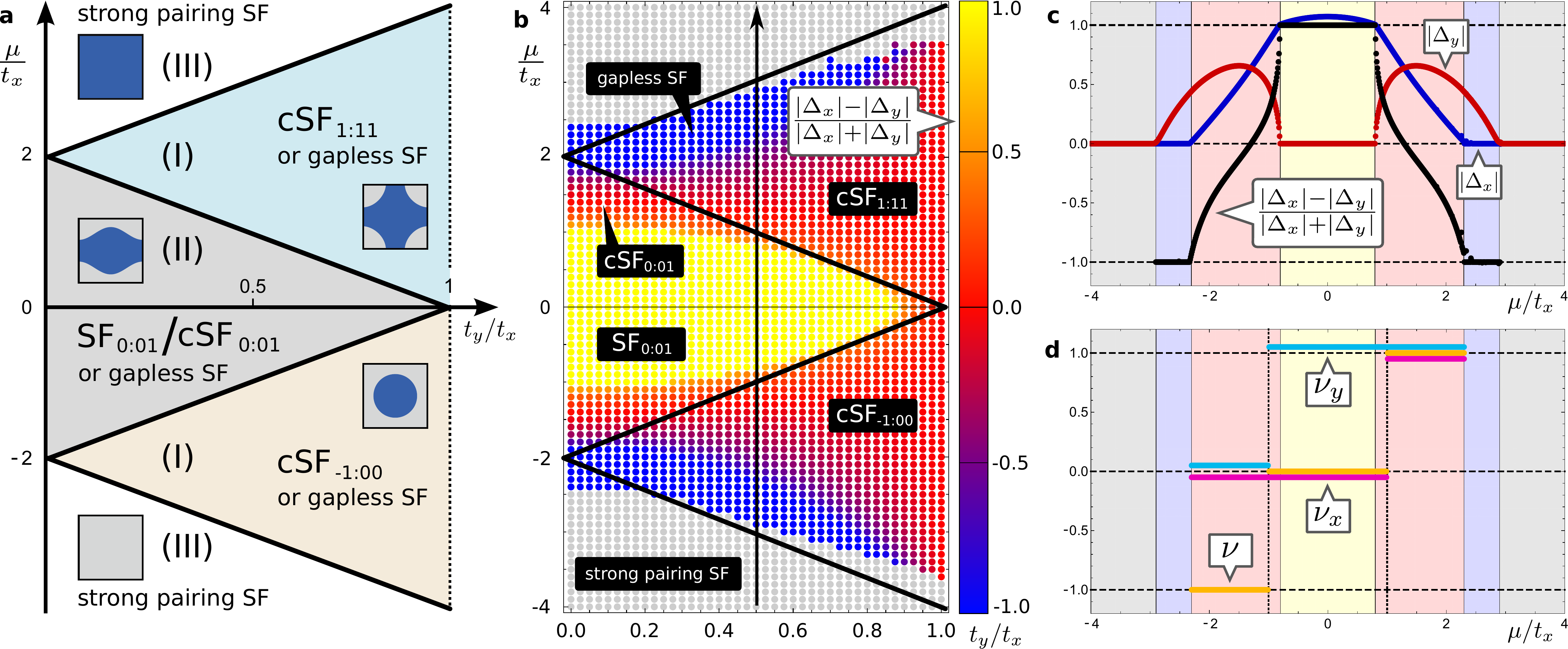}
  \caption{
    \textbf{Topological phases and  mean-field phase diagram.}
    \textbf{a}, Topological phase diagram: We can distinguish between three different topological regions.
    However, the topological indices also depend on the order parameter of the superfluid (see main text).
    The different  phases are denoted as SF$_{\nu:\nu_{x}\nu_{y}}$ for time-reversal invariant and 
    cSF$_{\nu:\nu_{x}\nu_{y}}$ for chiral superfluids with $\nu$ the strong topological index and $\nu_{x,y}$ the weak ones.
    \textbf{b}, Mean-field phase diagram for $\gamma t_{x}/g^2 = 1.5$: 
    For the rotationally symmetric setup with $t_{x}=t_{y}$ the ground state is given by a $p_x+i p_y$ superfluid. 
    While for strong anisotropy $t_{x} \neq t_{y}$ a pure $p_{x}$ or $p_{y}$ superfluid order parameter dominates. 
    The grey dots mark the points, where the gap parameters are too small for a convergence of the numerical calculations, 
    and therefore no superfluid phase is accessible for experimentally realistic temperatures.  
    \textbf{c-d} shows the gap parameters $\Delta_{x}$, $\Delta_{y}$,  and $(|\Delta_{x}|-|\Delta_{y}|)/(|\Delta_{x}|+|\Delta_{y}|)$ along a cut through the phase diagram 
    at $t_{y}=t_{x}/2$ [see arrow in \textbf{b}], as well as the different topological indices. It is important to stress, that the topological index $\nu_{y}$ 
    jumps to 1 at a different position, than the vanishing of $\Delta_{y}$, i.e., the topological transitions are essentially decoupled from the mean-field transitions.
  }\label{fig2}
\end{figure*}

\section{Mean-field theory}

We first study the zero temperature phase diagram within  mean-field theory.  Such a mean-field analysis is well justified as recent numerical calculations in a particle number conserving approach have  demonstrated the appearance of a $p$-wave superfluid exhibiting Majorana modes \cite{kraus13}. The authors of Ref.~\cite{kraus13} studied a two wire setup with a similar interaction between the wires which quantitatively agrees with the mean-field expectations. We tested  that the same  agreement is also valid for a three-wire setup.
As we are here interested in a higher dimensional setup, we expect that the quantitative behavior of the phase diagram is again well captured within mean-field theory. The resonant coupling to the molecular states gives rise to superconducting $p$-wave pairing via the formation of a Bose-Einstein condensate in the molecular states. Introducing the order parameters $\Delta_{x} = 4 g \langle  \sum_{p} X_{p}\rangle/ N $ and $\Delta_{y} = 4 g \langle  \sum_{p} Y_{p}\rangle/N $ for the macroscopic occupation of the molecular state at zero momentum, the Hamiltonian reduces to a quadratic fermionic theory
\begin{displaymath}
 H= \frac{1}{2 } \sum_{{\bf q}}\left( \begin{array}{c} c^{\dag}_{{\bf q}} \\ c_{-{\bf q}}
 \end{array} \right)^T \left(\begin{array} {c c}
 \epsilon_{{\bf q}} & \Delta_{\bf q}\\
  \Delta_{\bf q}^{*}& - \epsilon_{\bf q}
 \end{array}  \right) \left( \begin{array}{c} c_{{\bf q}} \\ c^{\dag}_{-{\bf q}} 
 \end{array} \right) + \mathcal{F}_{0}.
\end{displaymath}
Here, $c_{\bf q} = \sum_{i} e^{i {\bf q} {\bf r}_{i}} c_{i}/\sqrt{N}$, and $N$ denotes the number of lattice sites. 
$\mathcal{F}_{0}$ accounts for the conventional operator independent parts. Furthermore, the tight-binding dispersion  for the fermions reduces to 
$\epsilon_{\bf q} = - 2 \sum_{\alpha \in\{x,y\}} t_{\alpha} \cos(q_\alpha a)  - \mu$,
while the gap parameter  takes the form of a $p$-wave superfluid 
\begin{equation}
 \Delta_{\bf q} =  - i \left[  \Delta_{x} \sin(q_{x } a)+\Delta_{y} \sin(q_{y } a) \right],
\end{equation}
where $a$ denotes the lattice spacing. Using a Bogoliubov transformation, we obtain the superfluid excitation spectrum
$E_{\bf q} = \sqrt{\epsilon_{\bf q}^2 + |\Delta_{\bf q}|^2}$, and the ground-state energy per unit cell
\begin{equation}
  \mathcal{F}(\Delta_{x},\Delta_{y}) = \int \frac{\mathrm{d}{\bf q}}{v_{0}} \frac{ \epsilon_{\bf q} - E_{\bf q}}{2}    +  \frac{\gamma}{16 g^2} \Big[ |\Delta_{x}|^2 +|\Delta_{y}|^2 \Big],
\end{equation}
with $v_{0} = (2 \pi)^2/a^2$ denoting the volume of the first Brillouin zone. The order parameters $\Delta_{x}$ and $\Delta_{y}$ are determined by the gap equation minimizing the ground state energy
\begin{equation}
\partial_{\Delta_x} \mathcal{F}(\Delta_{x},\Delta_{y}) = \partial_{\Delta_y} \mathcal{F}(\Delta_{x},\Delta_{y}) =0.
\end{equation}
The results of the mean-field theory are shown in Fig.~\ref{fig2}b: we find a $p_{x} + i p_{y}$ superfluid for the fully isotropic setup with $t_{x} = t_{y}$ where $\Delta_{x} = \pm  i \Delta_{y}$. In addition to the $U(1)$ symmetry breaking, this phase also breaks time reversal symmetry. For finite interaction strength the $p_{x} + i p_{y}$ superfluid is stable to a small anisotropy in the hopping. Note that the anisotropic behavior is reflected in the order parameter, i.e. $|\Delta_{x}| \neq |\Delta_{y}|$.
However, we denote a $p_{x} + i p_{y}$ superfluid as a phase with a finite order parameter $\Delta_{x}$ and $\Delta_{y}$ obeying the fixed phase relation $\Delta_{x}/\Delta_{y} = \pm i |\Delta_{x}/\Delta_{y}|$.  For increasing anisotropy $t_{x}\neq t_{y}$ transitions into a $p_{x}$ ($p_{y}$) superfluid can appear, depending on the value of the chemical potential $\mu$.

\section{Topological phase transitions}

In addition to the mean-field transitions, the lattice system also exhibits a series of  topological quantum phase transitions beyond those found in the classification of continuum 2D superfluids \cite{ReadGreen00}.
 In the parameter regime, 
where the superfluid exhibits an excitation gap, the topological properties  are characterized by three topological indices \cite{daichi12}: the first denotes the strong topological index given by the Chern number $\nu$ characterizing the two-dimensional $p_{x}+ i p_{y}$ superfluid, and takes values $\nu = 0, \pm 1$, see Fig.~\ref{fig2}a. In addition, the system exhibits two weak topological indices \cite{Fu2007a,Fu2007},  which we denote as $\nu_{x} = 0,1$ and $\nu_{y}= 0,1$. The latter quantities are responsible for the appearance of 
Majorana modes in Kitaev's Majorana wire \cite{Kitaev:2001gb}, and can be finite in $p_{x}$ superfluids as well as chiral $p_{x}+i p_{y}$ superfluids. 
It is important to stress that the phase boundaries for the topological phase transitions are independent of the strength of the superfluid order parameters, and they only depend on the topology of the Fermi surface. Therefore, we can distinguish between three different regions, see Fig.~\ref{fig2}a:
region (I) with a closed Fermi surface, region (II) with an open Fermi surface, and finally the strong pairing regime (III), where in absence of interactions the system is in a trivial band insulating (vacuum) state.  In the latter region (III), the superfluids exhibit no topological order with $\nu=\nu_{x}=\nu_{y}=0$; thus it is not of interest in the following.

The combination of the topological indices with the superfluid order parameter allows us now to characterize the different  phases. We use the notation SF$_{\nu:\nu_{x}\nu_{y}}$ for time-reversal invariant superfluids and  cSF$_{\nu:\nu_{x}\nu_{y}}$ for chiral superfluids.
First, we start with the chiral $p_{x} + i p_{x}$ superfluid. Here, we obtain two 
fundamentally different  topological phases, see Fig.~\ref{fig2}a:
 (I)  the strong topological superfluids cSF$_{-1:00}$ and cSF$_{1:11}$ with a finite Chern number $\nu=\pm 1$.  Within the standard symmetry classification scheme \cite{Altland1997,Schnyder2008,Kitaev2009b}, the cSF$_{\pm1:\nu_{x}\nu_{y}}$
 phase is in the symmetry class D (particle-hole symmetry).
 It is a special property of this phase that the weak indices depend on the chemical potential, i.e., 
 we obtain $\nu_{x}=\nu_{y}=1$ for $\mu>0$ and $\nu_{x}=\nu_{y}=0$ for $\mu<0$. This property will strongly influence the Majorana modes, see below.
In region (II), we find a weak topological superfluid in the symmetry class D (cSF$_{0:01}$). On the other hand, for the  $p_{x}$ superfluid, we obtain a weak topological superfluid (SF$_{0:01}$) in region (II) which belongs to the class BDI, see Fig.~\ref{fig2}a. 
While in the region with closed Fermi surface (I) the superfluid phase becomes gapless without any topological properties. For completeness, we point out that the $p_{y}$ superfluid is gapless in region (I) and (II).

The full phase diagram is then obtained by combining the mean-field phase diagram  with the topological properties. Its details strongly depend on the strength of the coupling parameters. Here, we are mainly interested in strong couplings with $g^2/\gamma  \sim t_{x}, t_{y}$ with large superfluid gaps.   Most remarkably, we find that for $\gamma t_{x}/g^2 =1.5$ all of the above discussed topological phases are realized for varying values of $\mu/t_x$ and $t_{y}/t_{x}$, see Fig.~\ref{fig2}.

\begin{figure*}[t]
 \includegraphics[width= 1.0\linewidth]{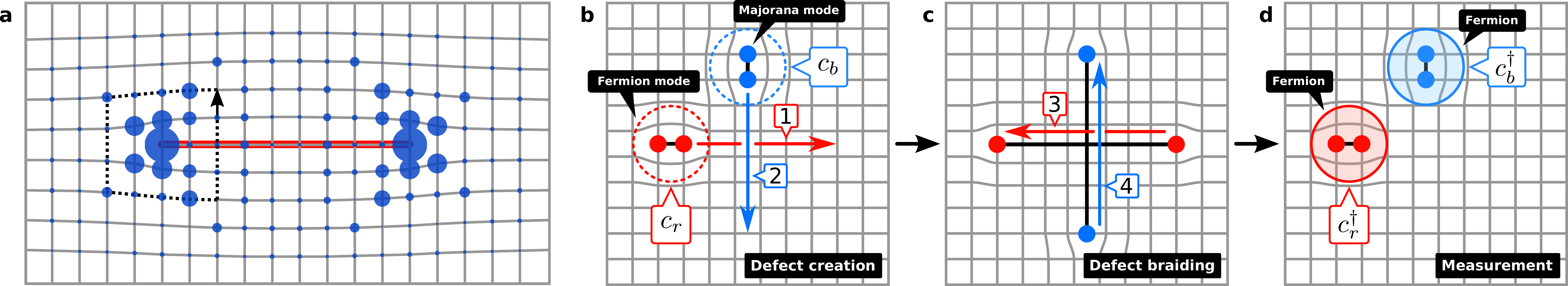}
  \caption{
    \textbf{Braiding of dislocation based Majorana modes as touchstone of non-Abelian statistics.}
    \textbf{a}, Majorana modes at lattice dislocations: an edge dislocation pair with Burgers 
    vector $\pm {\bf e}_{y}$ (black arrow) forming a single quantum wire (red line) immersed into the bulk
    superfluid. For a finite weak topological index $\nu_{y}=1$ such a setup generates two
    Majorana modes at the end of the wire. The localized wave function of the two Majorana 
    modes  (blue circles) is determined by numerically solving the Bogoliubov-de-Gennes equations.
    \textbf{b}, Four Majorana modes are generated by the formation of two dislocation pairs, and then spatially separated: first along path 1 and subsequently along path 2. 
    \textbf{c}, The braiding operation is achieved by recombining the Majorana modes along path 3 and then path 4.  
    \textbf{d}, The braiding transforms the initially unoccupied state at each dislocation pair into an occupied fermion mode. 
    The subsequent measurement of the unpaired fermions is a unique signature of the non-Abelian braiding statistics of the Majorana modes.
  }\label{fig5}
\end{figure*}

\section{Majorana modes at edge dislocations}

Associated with the topological index, we expect the appearance of Majorana modes at topological defects in the system.  Here, such topological defects can either be vortices or --- as a distinct feature of the lattice setup --- also lattice dislocations. Generally, we expect the Majorana modes in the vortex core for  Chern number $\nu=\pm 1$, while the weak index $\nu_{x,y}$ gives rise to Majorana modes localized at lattice dislocations with Burgers vector $\pm{\bf e}_{x,y}$. The latter can be easily understood in the limit $t_{y}=0$, where the system reduces to coupled one-dimensional wires: then, a pair of dislocations corresponds to the inclusion/removal of a one-dimensional wire of finite length into the bulk 2D system, see Fig.~\ref{fig5}a. This bulk superfluid induces a $p$-wave superfluid onto this single chain realizing the ideal toy model of a single Majorana chain \cite{Kitaev:2001gb};
 this behavior is in analogy to the proposals for the realisation of Majorana modes in solid state systems \cite{Sau:2010cl,Alicea:2010hy}.   
 
 The existence of Majorana modes is most conveniently verified using the Bogoliubov-de-Gennes equation, which can be efficiently solved numerically, an example is illustrated in Fig.~\ref{fig5}a. In summary, we find for the cSF$_{1:\nu_{x}\nu_{y}}$ phase  Majorana modes in the core of vortices. Most remarkably, the system also exhibits Majorana modes at edge dislocations  ${\bf e}_{x,y}$, but only for positive chemical potential $\mu> 0$ with a finite weak topological index $\nu_{x}=\nu_{y}=1$. Similarly, the cSF$_{0:01}$ and  SF$_{0:01}$  only exhibit Majorana modes at edge dislocations with Burgers vector $\pm{\bf e}_{y}$.  
Nevertheless, the two phases show distinct features due to their different symmetry classification: for the chiral phase cSF$_{0:01}$ in symmetry class D, the topological index is a $\mathbb{Z}_{2}$ index. Therefore, a pair of double dislocations with Burgers vector $\pm 2 {\bf e}_{x, y}$ leads to a hybridization of the Majorana modes, and consequently no ground state degeneracy. In turn, the time reversal symmetric phase  SF$_{0:01}$ is in
 the symmetry class BDI, which gives rise to a $\mathbb{Z}$ topological index. Consequently, a pair of double dislocations essentially describes a two wire setup and provides four Majorana modes with a four-fold ground state degeneracy. This behavior is well confirmed within the numerical solution of the Bogoliubov-de-Gennes equations.

\section{Braiding of non-Abelian anyons and Outlook}

In cold atomic gases, an edge dislocation corresponds to a vortex in the optical field generating the
optical lattice \cite{Schonbrun2006}. Such edge dislocations are most conveniently generated in a setup with local site addressability  
\cite{bakr09,sherson10,nogrette14}, where arbitrary shapes of the lattice can be achieved. In combination with
a time dependent modulation of the masks generating the lattice,  a full spatial and temporal control on edge dislocations 
is foreseeable in the near future. Such a setup then offers the opportunity for the observation of the non-Abelian statistics of Majorana modes by braiding the dislocations.  While the braiding of vortices in a superfluid has previously been predicted for the observation of the non-Abelian statistics \cite{tewari07,zhu11}, such experiments suffer from the difficulty to control a collective degree of freedom such as the superfluid phase, and the problem to insert adiabatically vortices into a superfluid. Here, edge dislocations in the lattice are much more favorable due to the precise and simple control on lattice structures available in cold atomic gases.

In the following, we present the protocol for measuring the non-Abelian statistics of the Majorana fermions, 
see Fig.~\ref{fig5}b-d. It is important that all operations are performed \textit{adiabatically}, i.e., slower than the characteristic time scale given by the superfluid gap. 

(a)  In a first step, we initialize the system by adiabatically creating two dislocation pairs. At each pair, we obtain a single fermionic mode described by the operators $c^{\dag}_{r,b}$  with a finite energy gap. This fermion mode is unoccupied as at low temperatures all fermions are Cooper paired.  The next step separates the two dislocation pairs first along path 1 and then along path 2. This operation splits the fermionic modes into  Majorana modes localized at the edge dislocations, and gives rise to a four-fold degenerate ground state of which two are accessible at fixed fermion number parity. However, adiabaticity of the process ensures a well defined initial state with $c_{r,b}|g\rangle=0$. 

(b) Next, we perform the braiding by  recombining the two dislocation pairs along path 3 and finally path 4. This process corresponds to moving the two Majorana modes around each other. According to the general non-Abelian braiding rules for Majorana modes \cite{Kitaev2006}, this transforms the fermionic operators via $c_{r,b} \rightarrow c^{\dag}_{r,b}$; here, we drop a phase factor, which is irrelevant for the protocol. As a consequence, the initially unoccupied state becomes occupied by one fermion each, i.e. $c^{\dag}_{r,b}|g\rangle =0$. 
In a physical picture, the braiding operation takes one Cooper pair from the superfluid condensate and splits it into two fermions with one residing at each dislocation pair.  

To probe the system one ramps the energy difference of the molecular state to the free fermionic states $\hbar \omega$ to negative values, which drives the system into the strong pairing phase with all paired fermions
 residing in the center of the plaquettes. This procedure is the analogue to the process of forming pairs via a Feshbach resonance \cite{winkler06}. Finally, a measurement of the fermionic density on the original lattice sites  \cite{bakr09} probes the unpaired fermions in the system. Here, we expect one unpaired fermion at each dislocation pair.  In order to test the protocol against induced noise, finite temperature, or violation of adiabaticity, one can test the process against a background measurement with a reversed order of path 3 and 4. Since this process does not braid the two Majorana modes, no unpaired fermions should be present in an ideal experiment.

{\bf Acknowledgements:}
We acknowledge support by the Center for Integrated Quantum Science and Technology (IQST) and the Deutsche Forschungsgemeinschaft (DFG) within SFB TRR 21, the Leverhulme Trust (ECF-2011-565), the Newton Trust of the University of Cambridge,  the Royal Society (UF120157), SFB FoQus (FWF Project No. F4006-N16), the ERC Synergy Grant UQUAM,  SIQS, and Swiss National Science Foundation. GM, SH, CK, and HB thank the Institut d'Etudes Scientifiques Carg\`ese and CECAM for their hospitality.

\section{Appendix}

{\bf Microscopic setup.} The fermionic states described by the operators $c_{i}$ ($c_{i}^{\dag}$) reside
on the sites of the optical lattice and are in the lowest Bloch band. The design of the interaction requires 
the coupling of these states to  different internal states trapped by an optical lattice with the minima in the 
center of the plaquettes. Such a setup is most conveniently achieved for cold atomic gases with a 
metastable $^{3}P_{2}$  state such as $^{87}\rm{Sr}$ or $^{171}\rm{Yb}$. Then, the 
metastable  $^{3}P_{2}$ states  are trapped at the sites of the lattice, while the ground 
state $^1S_{0}$ is trapped in the center  of the plaquette for an optical lattice close to the 
anti-magic wavelength. Therefore, the setup requires only a single two-dimensional optical lattice.
In addition, light assisted two-particle losses from the metastable $^{3}P_{2}$ are quenched due to
the fermionic statistic.

\begin{figure}[t]
 \includegraphics[width= 1\columnwidth]{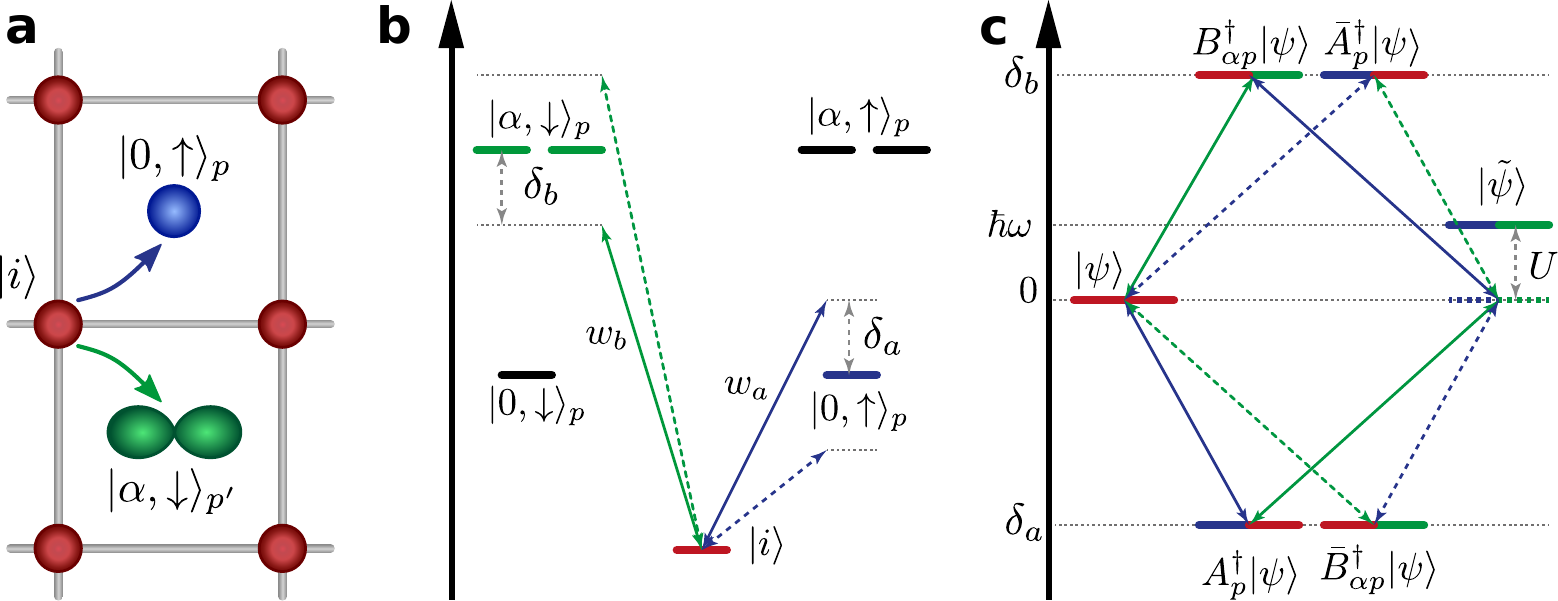}
  \caption{{\bf Microscopic setup.} {\bf a}, The lattice sites $|i\rangle$ are coupled to two different internal states trapped in the center of the plaquette.
  The first one $|0,\uparrow\rangle_{p}$ exhibits an $s$-wave orbital symmetry, while the second one  $|\alpha,\downarrow\rangle_{p}$
  shows a $p$-wave symmetry and is two-fold degenerate.  {\bf b},
  Single particle level structure with the relevant transitions (solid lines). The additional transitions (dashed lines) are required for the design of the desired coupling Hamiltonian.   {\bf c}, Energy levels for the two-particle states with the two interfering paths: $|\psi\rangle$ describes the state with two fermions on the lattice sites surrounding the plaquette, while $|\tilde \psi\rangle = (A^{\dag}_{p} B_{\alpha p}^{\dag}-\bar{A}^{\dag}_{p} \bar{B}_{\alpha p}^{\dag}) |\psi\rangle$  describes the near resonant  repulsively bound molecule with $p$-wave symmetry.}\label{fig4}
\end{figure}

Next, we focus on  the state trapped in the center of the plaquette. 
We are interested in two different hyperfine states in the electronic the ground state  $^1S_{0}$, which 
will be denoted by a spin index $\sigma$  
with $\sigma \in \{\downarrow, \uparrow\}$, and a setup with suppressed tunneling between different plaquettes.
The lowest lying state exhibits $s$-wave symmetry and will be denoted as $|0,\sigma\rangle_{p} $,
while the the  first excited state  $|\alpha,\sigma\rangle_{p}$ with  $\alpha\in \{x,y\}$ is two fold degenerate 
and exhibits a $p$-wave symmetry, see Fig.~\ref{fig4}b.

The coupling between the states on the lattice to the center of the plaquette is driven by two Raman transitions. The first Raman transition with detuning $\delta_{a}$ couples to the state $|0,\uparrow\rangle_{p}$ 
providing the Hamiltonain  $H_{a} =  w_{a}  \sum_{p}( A^{\dag}_{p}+ A_{p})$ with
$  A^{\dag}_{p} =   a^{\dag}_{p}\left(c_{1}+ c_{2} + c_{3} + c_{4}\right)$,
and the operator $a^{\dag}_{p}$ creating a fermion in the state $|0,\uparrow\rangle_{p}$.
The coupling strength  $w_{a}$ accounts for the Rabi frequency as well as the wave function overlap. 
 Note that the form of the coupling is determined by the $s$-wave symmetry of the state $|0,\uparrow\rangle_{p}$.
In analogy, the second Raman transition couples to the states $|\alpha,\downarrow\rangle_{p}$  with fermionic operators  $b^{\dag}_{\alpha p}$  ($\alpha\in\{x,y\}$) and the detuning $\delta_{b}$. In order to simplify the discussion, we set $\delta_{b}=-\delta_{a}$ and $w_{a}=w_{b}$. Then, the coupling Hamiltonian reduces to
$H_{b} =  w_{b}  \sum_{p \alpha}( B^{\dag}_{\alpha p}+ B_{\alpha p}) $
with $B_{x,y p}^{\dag} =   b^{\dag}_{x,y p}\left(c_{1}\pm c_{2} -c_{3}\mp c_{4}\right)$.
Note, the different couplings due to the orbital $p$-wave symmetry of the states $|\alpha,\downarrow\rangle_{p}$.

The main idea for the design of the interaction is now the fact, that  the state $a^{\dag}_{p} b^{\dag}_{\alpha p}|0\rangle$ with
two fermions in the center of the plaquette exhibits a strong onsite interaction $U$ due to the $s$-wave scattering 
between two different hyperfine states. Within the rotating frame its energy is given by $\hbar \omega=\delta_{a}+\delta_{b}+U=U$, see Fig.~\ref{fig4}c. This motivates the introduction of two bosonic molecular states $X^{\dag}_{p} = a^{\dag}_{p} b^{\dag}_{x p}$ and $Y^{\dag}_{p} = a^{\dag}_{p} b^{\dag}_{y p}$ exhibiting orbital $p$-wave symmetry. 
For a choice of the detunings with $\hbar |\omega| \ll |\delta_{a}|$, we can then adiabatically eliminate all states with a single fermion in the center of the plaquette and arrive at the effective coupling Hamiltonian
\begin{equation}
   H_{c} = \bar{g}\sum_{p, \alpha} \left[ B^{\dag}_{\alpha p}A_{p}^{\dag} +A_{p}  B_{\alpha p}\right] \, ,
\end{equation}
with $\bar{g}=  |w_{a}|^2 U/(U^2-\delta_{a}^2)$. 
Note, that we have omitted additional terms describing an induced hopping of the fermionic operators  $c_{i}$; these 
terms will be discussed below. The  resonant coupling of the fermionic states $c_{i}$ to the $p$-wave molecules $X_{p}$ and $Y_{p}$
residing in the center of the plaquette reduces to
\begin{eqnarray}
   B^{\dag}_{x p} A^{\dag}_{p}&= & 2 X^{\dag}_{p} \left[ c_{2} c_{3} - c_{4} c_{1} + c_{1}c_{3}+ c_{2} c_{4}\right],\\ 
   B^{\dag}_{y p} A^{\dag}_{p}&= & 2 Y^{\dag}_{p} \left[ c_{1} c_{2} - c_{3} c_{4} + c_{1}c_{3}- c_{2} c_{4}\right].
\end{eqnarray}
This coupling term differs from the desired interaction in Eq.~(\ref{molecularcoupling}); the last two terms, describe 
a second representation of the $p$-wave symmetry for the coupling. While this coupling Hamiltonian gives rise
to interesting  $p$-wave superfluids, it is desirable to suppress these additional coupling terms.

In the following, we present a scheme, which completely quenches these terms, while for an experimental realisation it is sufficient to 
weakly suppress them. The scheme is achieved by an additional  transition with opposite detunings but equal coupling strengths, where the phase is spatially varying. The main requirement on the phase is, that the coupling to the state $c_{1}$ exhibits the opposite sign than the coupling to $c_{2}$, while $c_{1}$ and $c_{3}$ have the same sign.  
The desired behaviour is achieved employing the principles of Ref.~\cite{Jaksch:2003ud}, by adding Raman lasers with a contribution of the wave vector $\mathbf{k}_{\parallel}$ within the plane of the optical lattice; i.e.,  $\mathbf{k}_{\parallel}= k_0 (\mathbf{e}_x - \mathbf{e}_y)$  with $k_0$ the wavelength of the square lattice potential. (By contrast, the Raman lasers for transitions $A_p^\dag$, $B_p^\dag$ must be incident at a right angle to the system.)
Then, we obtain the additional coupling terms $  \bar{A}^{\dag}_{p}  =   a^{\dag}_{p}\left(c_{1}- c_{2} +c_{3} - c_{4}\right)$
and 
$ \bar{B}_{x,y p}^{\dag} =   b^{\dag}_{x,y p}\left(c_{1} \mp c_{2} - c_{3} \pm c_{4}\right)$.
The full coupling Hamiltonian exhibits interference between the two independent excitation channels for the molecules, see Fig.~\ref{fig4}c, and reduces
to
\begin{equation}
   H_{c} =  \bar{g} \sum_{p, \alpha} \left[ B^{\dag}_{\alpha p}A_{p}^{\dag} -  \bar{B}^{\dag}_{\alpha p} \bar{A}^{\dag}_{p} + {\rm h.c.}\right], 
 \end{equation}
which reduces to the desired coupling in Eq.~(\ref{molecularcoupling}) with $\hbar \omega = U$ and $g=4 \bar{g}=4 |w_{a}|^2 U/(U^2-\delta_{a}^2)$. In addition, the induced hopping terms via the single excitation in the center of the plaquette reduces to an additional  conventional hopping as in Eq.~(\ref{hopping}) 
with $t_{x}=t_{y} = 2 |w_{a}|^2 /\delta_{a}$. Its interference with the direct hopping allows us to tune the ratios $g/U$ and $t/U$ independently.

\end{document}